\documentclass{PoS}
\PoS{PoS(LAT2005)124}
\usepackage{graphicx}
\newcommand{\srm}[1]{{\textrm{\scriptsize{#1}}}}
\newcommand{\DCI}{\ensuremath{D_\srm{CI}}}
\title{Implementing Hybrid Monte Carlo with stout-smeared chirally improved 
Dirac operators\footnote{for the Bern-Graz-Regensburg (BGR) collaboration}}

\ShortTitle{Implementing HMC with stout-smeared chirally improved Dirac operators}

\author{C. B. Lang\\
Karl-Franzens-Universit\"at Graz, Austria\\
E-mail: \email{christian.lang@uni-graz.at}}
\author{Pushan Majumdar\thanks{Supported by Fonds zur F\"orderung
der wissenschaftlichen Forschung in \"Osterreich, project M870-N08 (Lise Meitner
Fellowship)} \\

Karl-Franzens-Universit\"at Graz, Austria\\
E-mail: \email{pushan.majumdar@uni-graz.at}}

\author{\speaker{Wolfgang Ortner}\thanks{Supported by Fonds zur F\"orderung
der wissenschaftlichen Forschung in \"Osterreich, project P16310-N08}\\
Karl-Franzens-Universit\"at Graz, Austria\\
E-mail: \email{wolfgang.ortner@uni-graz.at}}

\abstract{We discuss our implementation of dynamical Ginsparg-Wilson type
fermions using a stout-smeared chirally improved Dirac operator. Such operators
have been studied extensively in quenched calculations within the
Bern-Graz-Regensburg (BGR)  collaboration. Here we discuss the development and
testing of the Hybrid Monte Carlo algorithm  with this Dirac operator. We study
the chiral properties of this operator in a dynamical setup, comparing, e.g.,
the spectra  of the operator for the dynamical and quenched cases. We then
discuss quantitative features of the algorithm like autocorrelation and
performance.}

\FullConference{XXIIIrd International Symposium on Lattice Field Theory\\

                 25-30 July 2005\\

                 Trinity College, Dublin, Ireland}

\begin{document}

\section{Chirally improved fermions}
Chirally improved (CI) fermions \cite{Ga01} are Ginsparg-Wilson type fermions.
The most general ansatz for a lattice Dirac operator reads
\begin{equation}\label{dcieq} 
 D_{x,y} = \sum_{i=1}^{16} \alpha^i_{x,y}(U)\Gamma_i \;,
\end{equation}
where $\Gamma_i$ $\{i=1\ldots 16\}$ are the 16 elements of the Clifford
algebra and  $\alpha^i_{x,y}(U)$ are sums of path ordered products of links $U$. Inserting $D$
into the  Ginsparg-Wilson relation and solving the resulting algebraic equations
yields the chirally improved Dirac operator $\DCI$.  In principle this can be an
exact solution, but that would require an infinite number of terms. 
In practice the number of terms is finite and the
operator is a truncated series solution to the Ginsparg-Wilson relation, respecting
the lattice symmetries, invariance under charge conjugation as well as 
$\gamma_5$-hermiticity, but connecting sites only over a certain distance.

\begin{figure}[th]
\begin{center}
  \includegraphics[width=8cm]{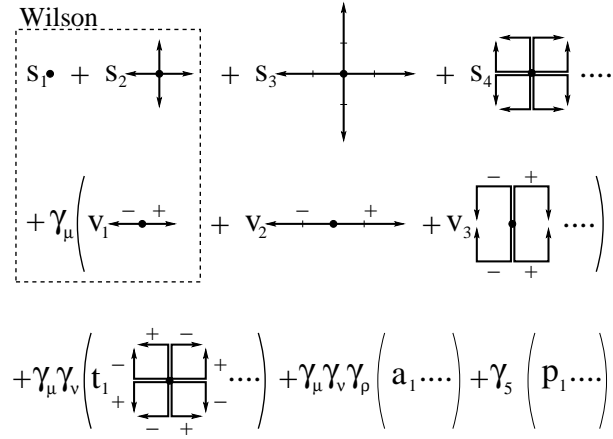}
  \caption{\label{fig:fig1}The ansatz for the chirally improved Dirac operator}
\end{center}
\end{figure}

For our implementation we use coefficients up to a path length of four
\cite{Ga01,GaHiLa00}.  Some of these terms used are sketched in
Fig.~\ref{fig:fig1}.
An operator like this, which is ultra-local,  can never fulfill the relation
exactly, which of the currently known actions only the overlap action does. 

CI fermions have been extensively tested in quenched calculations (see, e.g.,
Ref.~\cite{GaGoHa03a}). In these tests it was found that smearing the gauge
fields improves  the properties of CI fermions significantly and as a
consequence on smeared gauge configurations one can go to smaller quark masses
without running into the problem with exceptional configurations. On HYP smeared
quenched  configurations pion masses down to 220 MeV could be reached on
lattices of size \(16^3 \times 32\) and about 250 MeV on \(12^3 \times 24\) 
lattices. We expect this to improve in dynamical simulations. 

\section{Dynamical chirally improved fermions}
\subsection{First attempts}

First attempts to use CI fermions in dynamical simulations were presented in 
\cite{LaMaOr04}. As mentioned, smearing of the gauge fields is helpful for going
to small quark masses. In all corresponding quenched simulations up to now HYP
smearing was used. This, however, is not differentiable and therefore does not
allow for implementation in a  Hybrid Monte Carlo (HMC) updating algorithm. In
Ref.~\cite{LaMaOr04} we therefore tried to use a partial-global updating scheme
and we could collect first experiences on rather small lattices (up to 
\(8^4\)). 

The major drawback of such algorithms is that the computer time increases as
\(V^2\). We therefore decided to proceed with implementing the HMC method, which
is, however, technically  significantly more complicated for actions like ours,
that have many interaction terms. 

\begin{table}[t]
\begin{center}
  \begin{tabular}{rlllllllllll}
    \hline
    \hline
    \(L^3 \times T\)  & \(a\,m\) &$\beta_\srm{LW}$ & $res$     &$\Delta t$&steps& acc.       &HMC    & CPU Hours/   &\(a\)[fm]  \vspace{-3pt}\\
                      &          &                 &		 &	  &	&	     &time	   &HMC Time&		\\
    \hline
    \(8^3 \times 16\) &  0.05    & 5.4 & \(10^{-10}\) &0.015  & 50  &$\sim$93\%  &700      & 0.6 h (16 O)&  0.114(3)\\
    \(8^3 \times 16\) &  0.05    & 5.3 & \(10^{-10}\) &0.015  & 50  &$\sim$91\%  &700      & 0.6 h (16 O)&  0.135(3)\\
    \(8^3 \times 16\) &  0.08    & 5.4 & \(10^{-10}\) &0.015  & 50  &$\sim$93\%  &700      & 0.5 h (16 O)&  0.138(3)\\
    \(12^3 \times 24\)&  0.05    & 5.3 & \(10^{-8}\)   &0.01  & 100 &$\sim$87\%  &200      & 4 h (8 H)  &  0.129(3) \\
     \hline
  \end{tabular}
  \caption{\label{tab1}Parameters, statistics and some results from our
dynamical simulations.}
\end{center}
\end{table}

\subsection{Hybrid Monte Carlo}
In quenched calculations it was found that the tadpole improved L\"uscher-Weisz
gauge action had certain advantages over the Wilson gauge action in the sense
that the configurations produced with  this action were smoother than the ones
produced by the Wilson gauge action. We therefore used the same gauge  action in
our studies. 

Smearing is an essential quality improving ingredient in our Dirac operator. The
recent introduction of the differentiable ``stout''-smearing \cite{MoPe04} 
opened the possibility to implement HMC for smeared fermionic actions like ours.

In principle HMC is directly applicable for \ensuremath{D_\srm{CI}}; problems
like the discontinuities one has to deal with in HMC for the overlap action do
not occur for our action -- it is not an exact GW action, which is an advantage
in this regard.  It can  be expected that HMC for CI retains the very favorable
\(V^{5/4}\) scaling properties which it shows for staggered and Wilson fermions.

The problems of implementation were mostly technical. Much of the work went
into developing automated methods to write down the fermionic force for all 2428
different terms of our operator. Otherwise the HMC code we used so far is rather
conservative.  We use a standard pseudo\-fermion action with two mass degenerate
fermion flavors \(S_f = |D^{-1} \xi|^2 \). The integration scheme is leap-frog
integration, as the solver for \(D^\dagger D\) (needed for the molecular dynamics
evolution) the standard conjugate gradient algorithm, and as the solver of \(D\)
alone (needed in the accept/reject step) the BiCGStab was used.

It is well known that the performance of the inverter may be significantly
improved by providing a good guess for the solution.  Using an optimal linear
combination of the twelve previous solutions caused the
number of conjugate gradient iterations,  required for the inversion of the $\DCI$
during the molecular dynamics evolution inside a trajectory, to fall by a 
factor between two and three (chronological inverter by minimal residual
extrapolation).

Here we concentrate on the algorithmic aspects of our work; in another
contribution to these proceedings we give first results of the physical aspects \cite{LaMaOr05b}.

\subsection{Parameters and statistics}

First tests of our code were done on \(8^3 \times 16\) lattices. As we already
know from the quenched tests this lattice size does not allow for small bare
quark masses, we chose \(a\,m = 0.05\) and \(a\,m = 0.08\). Runs with \(12^3
\times 24\) lattices are in progress; there we approach much smaller masses
\(a\,m\) down to 0.02.

In Table~\ref{tab1} we present some parameters and results of our simulations.  The
conjugate gradient residual for the molecular dynamics evolutions is denoted as
$res$, initially, when we started our small lattice runs,  we chose a very small
value (\(10^{-10}\)). We then gradually increased it in accordance with the
folklore of other dynamical simulations (with other actions) and are currently
using $res=10^{-7}$ during the HMC trajectory but still \(10^{-10}\) in the
accept/reject step.

The lattice spacing \(a\)  was determined \cite{LaMaOr05b} using the Sommer
scale in order to have a first estimate. Eventually we will use the hadron
masses to set  the scale.

\begin{figure}[t]
\begin{center}
\includegraphics[width=10cm,clip]{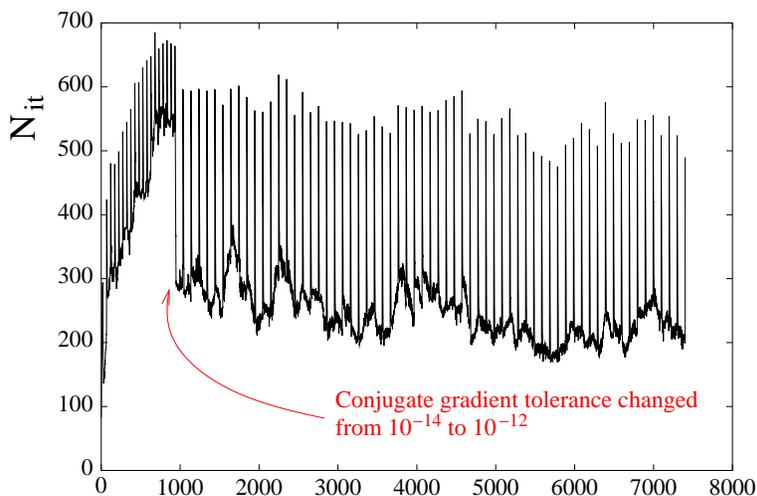}
\end{center}
\caption{\label{cgfig}Number of conjugate gradient
iterations for each molecular-dynamics step ($12^3\times 24$ lattice;  $a$=0.135 fm; $a\,m=0.05$). We use the
information from 12 previous solutions as starting guess for new solution.  The
peaks correspond to the number of iterations required in the accept/reject
step.}
\end{figure}

The runs were done on a cluster of Sun Fire V20z each with two Opteron 248 
processors at 2.2 GhZ (at the Karl-Franzens-Universit\"at, Graz) and on the
Hitachi  SR8000 at LRZ Munich. In the table (16 O) denotes 16 of the Opteron
processors (8 nodes) in parallel,  (8 H) denotes 8 SR8000 nodes. Just to get an
idea about timing, we note that  for the $12^3\times 24$ lattice at $a\,m=0.05$
one trajectory takes $\sim 4$ hours on 8 nodes of the  Hitachi.

As a first study we made runs on small lattices checking equilibration, lattice
spacing and the  the $\rho$ and $\pi$ meson correlators. 

\section{Results}
 
In Fig.~\ref{cgfig} we plot the number of conjugate gradient iterations  against
molecular dynamics time. After 1000 inversions we decided to reduce the
conjugate gradient tolerance was from $10^{-14}$ to $10^{-12}$. The drop in the
number of iterations is quite evident from this plot.

\begin{figure}[t]
\begin{center}
  \includegraphics*[width=8cm]{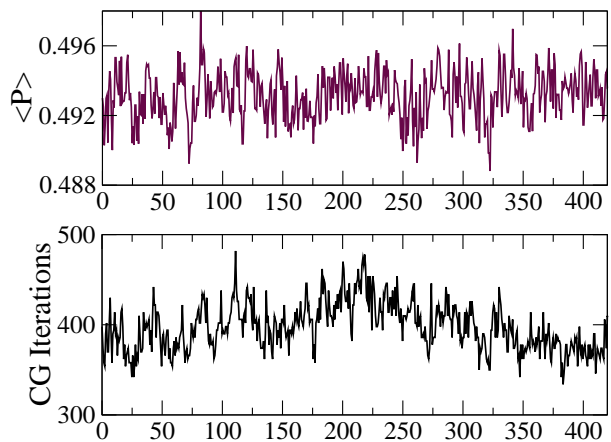}
\end{center}
\caption{\label{fig:autocorrhistory} History of the plaquette (upper part)  and
the number of CG iterations (lower part) in the accept/reject step (2nd run in
Table \protect \ref{tab1}). }
\end{figure}

To demonstrate autocorrelation we show in Fig.~\ref{fig:autocorrhistory} the
histories of the plaquette and the number of BiCGStab iterations in our
accept/reject step. Visual inspection indicates a time scale of roughly 25 and a
possible second time scale ${\cal O}(200)$. 

We computed the integrated  autocorrelation time from the number of BiCGStab
iterations required for the accept/reject step. This is widely believed to be
the longest autocorrelation present and in  our case this is about 36
trajectories. In contrast the autocorrelation time for the plaquette, which is
the most local quantity and therefore is expected to have the shortest
autocorrelation time is less than 2 trajectories.  Obviously the runs are yet to
short for a precise determination of the autocorrelation times.

\begin{figure}[ht]
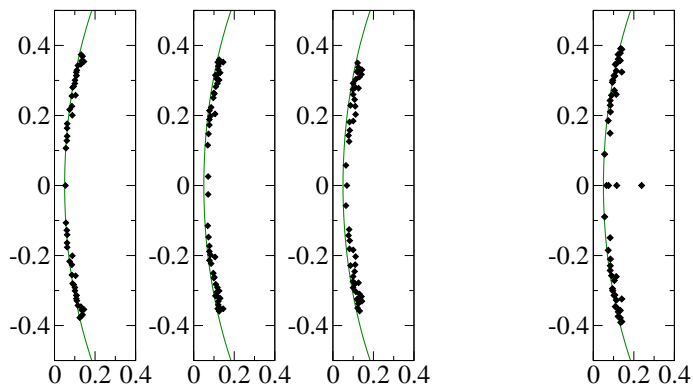

\begin{center}
\includegraphics[height=5cm,clip]{spectrum.eps}\qquad\qquad
\includegraphics[height=5cm,clip]{spec_quenched.eps}
\end{center}
\caption{\label{fig:spec}L.h.s: Three spectra of $\DCI$ (\(a\,m = 0.05\)) on
dynamical  configurations of \(8^3 \times 16, \ a = 0.14\) fm.  R.h.s: Sample $\DCI$
(\(a\,m = 0.05\)) spectrum on a quenched configuration  with similar parameters,
i.e., lattice size \(8^3 \times 16\), \(a = 0.15\) fm. }
\end{figure}

The $\DCI$ -- due to the truncation -- obeys the GW Relation only approximately. In
the quenched calculations it has exhibited, however, very good chiral
properties.  We may now check, whether this feature survives or even improves in
the simulation with dynamical fermions. As a simple check  we plot  in
Fig.~\ref{fig:spec} the low  lying spectrum of $\DCI$ along with the
Ginsparg-Wilson circle for three typical gauge configurations and compare with a
spectrum from a quenched simulation with similar lattice parameters. As can be
seen from the figure, the spectrum follows the Ginsparg-Wilson  circle
remarkably well at least for the low-lying modes, that have been determined, and
that are important towards the continuum limit.  This assures us to a certain
extent  that the good chiral properties observed in the quenched studies are
indeed also present in the dynamical case.
We conclude that the chiral properties of $\DCI$ on dynamical configurations are
roughly the same as on quenched configurations of the  same lattice spacing.

\section{Summary and outlook}

We presented recent results of our work in progress on implementing 
Ginsparg-Wilson type fermions in a full dynamical QCD simulation.  Dynamical
chirally improved fermions with stout-smeared links can be  implemented using an
HMC algorithm with all its well-known favorable properties, like \(V^{5/4}\)
scaling. The Dirac operator retains its good chiral properties  which allowed --
in quenched simulations -- to reach pion masses down to 250 MeV on moderately
sized lattice of  \(12^4 \times 24\). We expect to be able to reach such values
also in the dynamical fermion case. We hope to obtain pion  masses of about 300
MeV in our recently started runs and hope to be able to go below that in the
future. More results on our simulations, i.e., meson masses and topological
charge, can be found in \cite{LaMaOr05b}.


\end{document}